\documentclass[aip,graphicx]{revtex4}

\usepackage{graphicx}
\usepackage{dcolumn}
\usepackage{bm}
\usepackage{color}
\usepackage{amsmath}

\begin{document}

\title{Influence of non-universal effects on dynamical scaling in driven polymer translocation}

\author{T. Ikonen}
\affiliation{Department of Applied Physics and COMP Center of
Excellence, Aalto University School of Science,
P.O. Box 11000,
FI-00076 Aalto, Espoo, Finland}

\author{A. Bhattacharya}
\affiliation{Department of Physics, University of Central Florida, Orlando, Florida 32816-2385, USA}

\author{T. Ala-Nissila}
\affiliation{Department of Applied Physics and COMP Center of
Excellence, Aalto University School of Science,
P.O. Box 11000,
FI-00076 Aalto, Espoo, Finland}
\affiliation{Department of Physics,
Box 1843, Brown University, Providence, Rhode Island 02912-1843, USA}

\author{W. Sung}
\affiliation{Department of Physics, Pohang University of Science and Technology, Pohang 790-784, South Korea}

\date{\today}

\begin{abstract}
We study the dynamics of driven polymer translocation using both molecular dynamics (MD) simulations and a theoretical model based on the non-equilibrium tension propagation on the {\it cis} side subchain. We present theoretical and numerical evidence that the non-universal behavior observed in experiments and simulations are due to finite chain length effects that persist well beyond the relevant experimental and simulation regimes. In particular, we consider the influence of the pore-polymer interactions and show that they give a major contribution to the non-universal effects. In addition, we present comparisons between the theory and MD simulations for several quantities, showing extremely good agreement in the relevant parameter regimes. Finally, we discuss the potential limitations of the present theories.
\end{abstract}

\maketitle

\section{Introduction}
\label{sec:introduction}

The translocation of a polymer through a nano-sized pore occurs as a part of many biological processes and functions, such as DNA and RNA translocation through nuclear pores, protein transport across membrane channels and virus injection~\cite{albertsbook}. The translocation process is also envisioned to have several biotechnological applications, including rapid DNA sequencing, gene therapy and controlled drug delivery~\cite{meller2003}. Recently, the hope of realizing a third generation DNA sequencing method using nanopore sequencing devices has prompted rapid advancement in experimental methods and in technological applications~\cite{schadt2010, branton2008}, leading to commercialization of the nanopore sequencing technology in the near future~\cite{oxfordnanopore}.

However, despite the technological advances and considerable experimental~\cite{meller2003, kasi1996, storm2005}  and theoretical~\cite{chuang2001,kantor2004, sung1996, metzler2003, muthu1999, dubbeldam2007,vocks2008,sakaue2007,sakaue2008,sakaue2010,saito2011, rowghanian2011,luo2008,luo2009,milchev2011,huopaniemi2006,bhatta2009,bhatta2010,metzler2010,huopaniemi2007,lehtola2008,lehtola2009,gauthier2008a,gauthier2008b} efforts to understand the basic process, until very recently the fundamental physics of driven polymer translocation has remained elusive. The first attempts to model polymer translocation theoretically were made by Sung and Park~\cite{sung1996} and later by Muthukumar~\cite{muthu1999}, who viewed polymer translocation as a one-dimensional barrier crossing problem of the translocation coordinate $s$ (the length of the subchain on the {\it trans} side), with the activation barrier determined by the free energy of the polymer derived from equilibrium enumeration of random walks. However, if the process is facilitated by an external bias, the process is inherently out of equilibrium due to the long relaxation time of the polymer chain. This fact was first recognized by Kantor and Kardar~\cite{kantor2004}, and was later studied extensively in simulations~\cite{kantor2004,bhatta2010,bhatta2009,lehtola2009,luo2009}, with chain lengths and driving forces in the typical experimental regime. It was observed that with increasing driving force, significant non-equilibrium effects come into play~\cite{luo2009,bhatta2010}. The driven translocation problem was also studied by several authors with different approaches, such as the fractional Fokker-Planck method by Metzler and Klafter~\cite{metzler2003}, scaling theory supplemented with the fractional Fokker-Planck method by Dubbeldam et al.~\cite{dubbeldam2007} and  by Vocks et al. using a method based on local memory effects near the pore~\cite{vocks2008}. However, the first truly non-equilibrium treatment was given by Sakaue, using an ingenious tension propagation theory based on force balance, mass conservation and self-similarity of the polymer~\cite{sakaue2007,sakaue2008,sakaue2010,saito2011}. However, the theory was solved in the asymptotic limit of long chains, neglecting, e.g., the interaction between the pore and the polymer. It was pointed out already by Storm et al.~\cite{storm2005} that this interaction may significantly alter the dynamics of driven translocation, which was further demonstrated in simulations~\cite{lehtola2010,bhatta2010b,luo2007prl,luo2008pre}. Hence, the agreement of the theories with experiments and simulations remained modest.

Recently~\cite{ikonen2011}, we have generalized Sakaue's tension propagation (TP) model of driven translocation for finite chain lengths and included the pore-polymer interaction as an additional friction force. It was shown that the model is in excellent agreement with high-accuracy molecular dynamics simulations. In addition, we showed that due to the pore friction and finite size corrections to the tension propagation equations, the asymptotic limit is well beyond the present computational capabilities and simulation algorithms. For example, the scaling exponent $\alpha$, which relates the mean translocation time $\tau$ to the chain length $N_0$ as $\tau\sim N_0^\alpha$, actually retains a fairly strong dependence on the chain length even up to $N_0\approx 10^5$. Therefore, the scatter (non-universality) of $\alpha$, widely studied and reported in the literature, is in fact a finite chain length effect. 

The purpose of this work is to study further the non-universalities in driven polymer translocation brought on by finite chain length effects, and to see how the asymptotic limit is approached in different regimes. To this end, we use a combination of theoretical methods and molecular dynamics (MD) simulations. For the former, we use the theoretical Brownian dynamics tension propagation (BDTP) model introduced in Ref.~\cite{ikonen2011}, which is based on the tension propagation description and incorporates the finite size effects, and for the latter, we use full $N_0$-particle Langevin thermostatted MD simulations. The BDTP model and its numerical solution is described in Section~\ref{sec:model}, while the details of the MD simulations are included in Appendix A.

The central assumption of the tension propagation theory is that the effect of the {\it trans} side subchain on the non-equilibrium dynamics is small, and therefore its contribution to the effective total friction is neglected. This approximation is very good in the parameter regime typically used in experiments and simulations. However, in certain cases, the influence of the {\it trans} side chain on the dynamics may be non-negligible. Therefore, we do not expect perfect agreement with the BDTP model and the MD simulations. The limitations of the model are discussed in Section~\ref{sec:results}.  However, it will be shown that for most relevant situations, the BDTP model gives extremely good match with MD. In addition, it allows one to go beyond the chain length regime available to MD simulations ($N_0\approx 10^3$ for high driving force), and to see how the finite chain length effects diminish as the asymptotic limit is approached.  In addition, in Section~\ref{sec:results}, we present detailed comparison between the theory and MD simulations for the time evolution of the translocation coordinate $s(t)$, the scaling of translocation time with the driving force and the monomer waiting time distribution. In our previous work we have explained and demonstrated the applicability of the BDTP model to both two and three dimensional systems~\cite{ikonen2011}. In this work we provide new results for three dimensional geometry, some of which are directly relevant to gain better understanding of the experimental results for DNA translocation through nanopore. We also relate our findings to previous MD simulations found in the literature, and to the theories of Refs.~\cite{sakaue2007,sakaue2008,sakaue2010,saito2011,dubbeldam2011,rowghanian2011}.

\section{Model}
\label{sec:model}

\subsection{General framework}
\label{sec:model_general}

The purpose of the BDTP model is to present a coarse-grained, minimal model of driven polymer translocation~\cite{ikonen2011}. Since the driven translocation is a complex, non-equilibrium dynamical process, the rigorous solution of the full problem from first principles seems impossible at present. Instead, the BDTP model presents a phenomenological description, which interpolates between the low-force and high-force regimes. In the low-force regime, the fluctuations and the chain entropy become important. Hence, as a general framework, we adopt the description used by Sung and Park~\cite{sung1996} and Muthukumar~\cite{muthu1999}, the one-dimensional barrier crossing problem of the translocation coordinate $s$. Here, the chain starts from the {\it cis} side with one end inside the pore and is considered as translocated once $s=aN_0$, with $a$ the segment length. The free-energy due to chain entropy and the chemical potential difference $\Delta\mu$ is
$\mathcal{F}(s)=(1-\gamma')k_BT\ln\left[\frac{s}{a}\left(N_0-\frac{s}{a}\right)\right] +\frac{s}{a}\Delta\mu.$
Here $\gamma'$ is the surface exponent ($\gamma'=0.5,~\approx 0.69,~\approx 0.95$ for an ideal chain, and a self-avoiding chain in 2D and 3D, respectively), and $k_BT$ is the thermal energy. From $\mathcal{F}(s)$, the Brownian dynamics equation for $s$ in the overdamped limit follows as
$\Gamma\frac{ds}{dt}=(1-\gamma')k_BT\left[ \frac{1}{aN_0-s} -\frac{1}{s} \right] - \frac{\Delta\mu}{a} + \zeta(t).$
Here $\Gamma$ is the effective friction, and $\zeta(t)$ is Gaussian white noise satisfying $\langle \zeta(t) \rangle=0$ and $\langle \zeta(t)\zeta(t') \rangle = 2\Gamma k_BT\delta(t-t')$. In this framework, the non-equilibrium memory effects at larger driving forces are then taken into account by allowing the effective friction to depend on time, $\Gamma=\Gamma(t)$. This time-dependence is then solved from the tension propagation (TP) formalism.

\begin{figure*}
\centering
\includegraphics[bb=0.0cm 0.0cm 16.7cm 2.9cm]{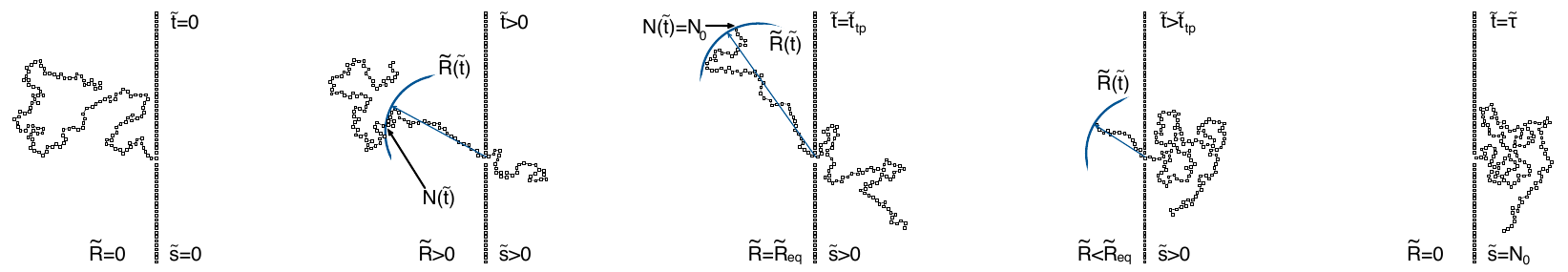}
\caption{(Color online) The time evolution of the polymer configuration during the translocation process, with time advancing from left to right. The arc denotes the position of the tension front $\tilde{R}$, which separates the chain into the moving and nonmoving domains. The last monomer inside the front is denoted by $N$ and the number of translocated monomers by $\tilde{s}$.
}
\label{fig:configuration}
\end{figure*}

The central idea of the tension propagation theory is to divide the subchain on the {\it cis} side into two distinct domains~\cite{sakaue2007,sakaue2008,sakaue2010,saito2011}. The first domain, closer to the pore, consists of all the monomers that are pulled towards the pore by the external driving force. The second domain consists of the remaining monomers, which are at rest (on the average).  As the driving force is applied at the pore, the chain begins to move in stages, with the segments closest to the pore being set into motion first. A close analogue is a coil of rope pulled from one end, which first has to uncoil and become tense before starting to move as a whole. To keep track of the moving part of the chain, one defines a {\it tension front}, which divides the chain into the moving and nonmoving domains. The front propagates in time as parts of the chain further away from the pore are set in motion, as shown in Fig.~\ref{fig:configuration}. As the monomers enter the front and start to move, the effective friction $\tilde{\Gamma}$ increases due to the increased drag between the polymer and the solvent. After a certain tension propagation time, $\tilde{t}_\mathrm{tp}$, the front reaches the end of the chain and the tension propagation process stops. After this time, the chain as a whole is pulled towards the pore. During this stage, the overall length of the subchain on the {\it cis} side decreases, which reduces the effective friction $\tilde{\Gamma}$. This stage continues until the last monomer reaches the pore and finally translocates at time $\tilde{t}=\tilde{\tau}$.

\subsection{Coarse-grained equations of motion}

In the BDTP model, the translocation process is described using only two degrees of freedom. The first one, the translocation coordinate $s$, describes the chain's position with respect to the pore, while the second one, the location of the tension front $R$, describes the response of the {\it cis} side chain to the driving force. To both of these degrees of freedom one has a corresponding equation of motion, which are coupled via the effective friction $\Gamma$. In the extremely high driving force limit, the location of the tension front can be described by only one equation, while in the more general case considered in this work, one needs to solve a system of several equations. The derivation of the BDTP model is presented in Ref.~\cite{ikonen2011} and the tension propagation theory in Refs.~\cite{sakaue2007,sakaue2008,sakaue2010,saito2011}. Therefore, in this work we do not reiterate the full derivation of the model, but rather present the resulting equations and outline the method of their numerical solution.

For convenience, we first introduce dimensionless units  denoted by the tilde symbol as $\tilde{X}\equiv X/X_u$, with the unit of length $a_u\equiv a$, force $f_u\equiv k_BT/a$, time $t_u\equiv \eta a^2 /k_BT$, velocity $v_u\equiv a/t_u$ and friction $\eta_u\equiv \eta$, where $\eta$ is the solvent friction per monomer. In these units, the Brownian dynamics equation for $\tilde{s}=s/a$ reads
\begin{equation}
\tilde{\Gamma}(\tilde{t})\frac{d\tilde{s}}{d\tilde{t}}=(1-\gamma')\left[ \frac{1}{N_0-\tilde{s}} -\frac{1}{\tilde{s}} \right] + \tilde{f} + \tilde{\zeta}(\tilde{t}) \equiv \tilde{f}_\mathrm{tot}, \label{eq:motion_dimless}
\end{equation}
where $\tilde{\Gamma}\equiv \Gamma/\eta$ and, for simplicity, we have assumed that the pore length $l_p=a$. Generalization of Eq.~(\ref{eq:motion_dimless}) to different pore lengths is straightforward (see, e.g., Ref.~\cite{gauthier2008a} for a similar case). The dynamics of Eq.~(\ref{eq:motion_dimless}) is essentially determined by  the effective friction $\tilde{\Gamma}$, which therefore must comprise all the dominant contributions of the non-equilibrium dynamics of the full polymer chain. Although under sufficiently large $\tilde{f}$, also the {\it trans} side subchain assumes a highly out-of-equilibrium configuration, it has been shown in Ref.~\cite{ikonen2011}, that for most cases the influence of the {\it trans} side chain on the dynamics is extremely small. This matter will also be further elucidated in Section~\ref{sec:results} of this work. Therefore, to a very good approximation it is sufficient to consider the drag force due to the {\it cis} side subchain and the frictional interaction between the pore and the polymer. Although the latter is negligible for infinitely long chains, it can significantly change the dynamics for finite chains. Formally, we can write $\tilde{\Gamma}$ as the sum of the {\it cis} side subchain and pore frictions, $\tilde{\Gamma}(\tilde{t})=\tilde{\eta}_\mathrm{cis}(\tilde{t})+\tilde{\eta}_p$. The first contribution can be calculated from the tension propagation formalism, while the pore friction $\tilde{\eta}_p$ has to be determined empirically. In the special case of extremely large driving force, one can find $\tilde{\eta}_\mathrm{cis}$ explicitly from the TP equations~\cite{ikonen2011}. More generally, however, it is easier to derive the velocity of the monomers at the pore entrance, $\tilde{v}_0$. In such a case, the effective friction is defined as
\begin{equation}
\tilde{\Gamma}(\tilde{t})= \frac{\tilde{f}_\mathrm{tot}}{\tilde{\sigma}_0(\tilde{t})\tilde{v}_0(\tilde{t})}, \label{eq:effective_friction}
\end{equation}
where $\tilde{\sigma}_0$ is the line density of monomers near the pore and $\tilde{\sigma}_0\tilde{v}_0\equiv d\tilde{s}/d\tilde{t}$ is the flux of monomers through the pore entrance. In either case, determining $\tilde{\Gamma}(\tilde{t})$ essentially reduces to calculating the number of moving monomers, whose combined drag force then constitutes the time-dependent part of the friction. 

For moderate driving forces (to be defined rigorously below) the chain on the {\it cis} side assumes a trumpet-like shape due to the higher stretching close to the point of action of the driving force. In this so-called trumpet (TR) regime, the line density near the pore is $\tilde{\sigma}_0=\tilde{f}_0^{1-1/\nu}$, where $\tilde{f}_0\approx \tilde{f}_\mathrm{tot}-\tilde{\eta}_p\tilde{v}_0$ is the force at the pore entrance~\cite{approx}. To solve the velocity $\tilde{v}_0$, one needs to solve a system of  four equations:
\begin{equation}
\frac{d\tilde{R}(\tilde{t})}{d\tilde{t}}=\tilde{v}_{R}(\tilde{t})  \left[ \frac{1}{\nu} A_\nu^{-1/\nu} \tilde{\sigma}_{R}(\tilde{t})^{-1} \tilde{R}(\tilde{t})^{1/\nu -1} \right]^{-1},  \label{eq:motionR}
\end{equation}
\begin{equation}
\tilde{\sigma}_R(\tilde{t})^{1/(1-\nu)}= \frac{\tilde{v}_0(\tilde{t})\tilde{R}(\tilde{t})}{\nu b \tanh(b)}\ln\left[\cosh\left(b\frac{\tilde{\sigma}_R(\tilde{t})^{\nu/(1-\nu)}}{\tilde{R}(\tilde{t})}\right)\right], \label{eq:xiR}
\end{equation}
\begin{equation}
\tilde{v}_R(\tilde{t})=\tilde{v}_0(\tilde{t})\frac{\tanh \left(b \tilde{\sigma}_R(\tilde{t})^{\nu/(1-\nu)} /\tilde{R} \right) }{\tanh(b)}, \label{eq:vR}
\end{equation}
\begin{equation}
\tilde{v}_0(\tilde{t})\tilde{R}(\tilde{t})C_b=\nu \tilde{f}_0^{1/\nu}, \label{eq:v0_tr}
\end{equation}
where $\nu$ is the Flory exponent, $A_\nu$ relates the chain end-to-end distance to its contour length as $\tilde{R}_{ee}=A_\nu N^\nu$. From MD simulations we have measured $A_\nu\approx 1.15\pm 0.03$ in three dimensions. The coefficient $C_b\equiv\frac{\ln [\cosh (b)]}{b\tanh(b)}$ with $b$ a dimensionless parameter that describes the velocity profile on the {\it cis} side and ensures global conservation of mass~\cite{ikonen2011}. The line density and velocity at the tension front are given by $\tilde{\sigma}_R$ and $\tilde{v}_R$, respectively. The numerical solution of Eqs.~(\ref{eq:motionR})--(\ref{eq:v0_tr}) is described in Section~\ref{sec:numerical}.

Equations~(\ref{eq:motionR})--(\ref{eq:v0_tr}) are used when the line density of the monomers at the pore entrance is greater than unity, i.e., for $\tilde{f}_0<1$. For slightly higher forces, the chain assumes a shape consisting of a fully elongated stem followed by a trumpet-shaped flower. In this stem-flower (SF) regime, the velocity at the pore entrance is given by
\begin{equation}
\tilde{v}_0(\tilde{t})\tilde{R}(\tilde{t})C_b=\tilde{f}_0 + \nu -1. \label{eq:v0_sf}
\end{equation}
Also in this regime, the line density and the velocity at the boundary are given by Eqs. (\ref{eq:xiR}) and (\ref{eq:vR}), respectively, and the time evolution of the front by Eq. (\ref{eq:motionR}). Note that Eqs. (\ref{eq:v0_tr}) and (\ref{eq:v0_sf}) ensure a smooth cross-over between the TR and SF regimes at $\tilde{f}_0=1$. In practice, we solve Eqs.~(\ref{eq:motion_dimless})--(\ref{eq:v0_sf}), choosing Eq.~(\ref{eq:v0_tr}) over Eq.~(\ref{eq:v0_sf}) if $\tilde{f}_0 < 1$, and vice versa.

In deriving Eqs.~(\ref{eq:xiR})--(\ref{eq:v0_sf}) we have adopted the form $\tilde{v}(\tilde{x},\tilde{t})=\tilde{v}_0(\tilde{t}) \frac{\tanh \left[ b\left( \tilde{x}/\tilde{R}+1\right)\right]}{\tanh(b)}$ for the velocity profile of the {\it cis} side subchain. Here $b$ is a dimensionless parameter that controls the sharpness of the profile and is fixed by enforcing global conservation of monomers, i.e., requiring that at the time of translocation $\tilde{s}(\tilde{\tau})=N_0$ and $\tilde{R}(\tilde{\tau})=0$. Although the adopted profile is a good approximation to MD simulations (see Appendix B), the exact functional form of the profile is not crucial. The essential requirements are that the profile is non-constant, goes to zero at the tension front ($\tilde{x}=-\tilde{R}$), and allows the global conservation of mass.

\subsection{Numerical solution of the BDTP equations}
\label{sec:numerical}

To solve the model numerically, we first set the initial values for the translocation coordinate $\tilde{s}$ and the location of the tension front $\tilde{R}$. In the initial configuration, the first chain segment is through the pore entrance and $\tilde{s}(\tilde{t}_0)=1$. This gives the initial condition for Eq.~(\ref{eq:motion_dimless}). Correspondingly, the initial location of the tension front is given by $\tilde{R}(\tilde{t}_0)=1/\tilde{f}_0$ for $\tilde{f}_0<1$ and $\tilde{R}(\tilde{t}_0)=1$ otherwise. This gives the initial condition for Eq.~(\ref{eq:motionR}). 

At the beginning of each time step, the random force $\tilde{\zeta}(\tilde{t})$ is sampled from a Gaussian distribution with the mean and variance given in Section~\ref{sec:model_general}. Since the value of $\tilde{s}$ is known at the start of the time step, the value of the total force $\tilde{f}_\mathrm{tot}$ is then also known. The task is then to determine the effective friction $\tilde{\Gamma}$ for the next time step. For this, we need to find the value for the velocity $\tilde{v}_0$ and the line density $\tilde{\sigma}_0$ [cf. Eq.~(\ref{eq:effective_friction})]. The velocity $\tilde{v}_0(\tilde{t})$ is given by Eq.~(\ref{eq:v0_tr}) in the TR regime, and by Eq.~(\ref{eq:v0_sf}) in the SF regime. The latter can be solved explicitly for $\tilde{v}_0$, while the former is solved numerically by the Newton-Raphson method~\cite{numerical_recipes}. Knowing $\tilde{v}_0$ then gives $\tilde{f}_0=\tilde{f}_\mathrm{tot}-\tilde{\eta}_p\tilde{v}_0$ and $\tilde{\sigma}_0=\tilde{f}_0^{1-1/\nu}$. The effective friction is then given by Eq.~(\ref{eq:effective_friction}), and Eq.~(\ref{eq:motion_dimless}) can be advanced in time by the explicit Euler algorithm~\cite{kloeden-platen}.

To solve the effective friction for the subsequent time steps, one also has to find the time evolution of $\tilde{R}$ from Eq.~(\ref{eq:motionR}). Therefore, one has to know the velocity $\tilde{v}_R$ and the line density $\tilde{\sigma}_R$ near the tension front. The line density is $\tilde{\sigma}_R=\tilde{\xi}_R^{1/\nu -1}$, with $\tilde{\xi}_R$ solved from Eq.~(\ref{eq:xiR}) by Newton-Raphson iteration. The velocity $\tilde{v}_R$ is then given by Eq.~(\ref{eq:vR}) and, with both $\tilde{v}_R$ and $\tilde{\sigma}_R$ known, the location of the tension front $\tilde{R}$ for the next time step can be solved from Eq.~(\ref{eq:motionR}).

We emphasize that the BDTP model is not an alternative formulation of molecular dynamics. It is a model of two degrees of freedom, $\tilde{s}(\tilde{t})$ and $\tilde{R}(\tilde{t})$. In addition, it has no free parameters that could be used to adjust the results. Therefore, the agreement of the model with MD simulations presented in both Ref.~\cite{ikonen2011} and in Section~\ref{sec:results} of this work is not self-evident, but a demonstration of the tension propagation formalism's capability to capture the essential physics of driven polymer translocation.

\subsection{Determining the pore friction $\eta_p$}

To complete the BDTP model, we still need to determine the pore friction $\eta_p$. Because $\eta_p$ characterizes the interactions between the pore and the polymer chain, we do not expect it to have any universal value. Rather, it should depend on the geometry and dimensions of the pore and, in a more refined model, on the chemical details of the pore and its immediate vicinity. In the context of this work, we only consider coarse grained models of the pore, where the pore consists of immobile monomers placed in a configuration that forms the edges of the pore (cf. Appendix A). The same approach has been used widely in the literature for MD simulations of polymer translocation~\cite{huopaniemi2006,bhatta2009,bhatta2010,luo2008,luo2009,metzler2010,dubbeldam2011}. While it is possible to determine the pore friction also from experimental data, in this work we only consider the numerical pores used in our benchmark MD simulations.
 
To determine $\eta_p$, we look at the movement of the first few monomers and fix $\eta_p$ in the BDTP model so that the velocity matches MD simulations. In this early stage of translocation, where the tension front $\tilde{R}$ is still close to the pore, the friction in the system is largely determined by the friction between the pore and the polymer, giving the most accurate estimate for $\eta_p$. In practice, we look at the waiting time per monomer $w(\tilde{s})$, defined as the time that the individual monomer spends inside the pore. With $\tilde{f}$ sufficiently large, $\tilde{w}\propto \tilde{\Gamma}/\tilde{f}$. For small $\tilde{s}$, the friction $\tilde{\Gamma}$ and $\tilde{w}$ are mostly determined by $\tilde{\eta}_p$. In Ref.~\cite{ikonen2011}, we have measured $\eta_p$ specifically for the pore geometries used in Refs.~\cite{huopaniemi2006,luo2009}, for which we had access to the waiting time distributions. In these cases, we have $\eta_p\approx 5$ and $\eta_p\approx 4$ in 3D and 2D, respectively. Other pore geometries and the effect of $\eta_p$ on the translocation dynamics will be discussed further in Sec.~\ref{sec:results}. Finally, it should be noted that for each pore geometry,  $\eta_p$ is fitted only once, as opposed to being done separately for each combination of $\tilde{f},\eta$, etc. Thus, $\eta_p$ is not a freely adjustable parameter, but is a property of the pore.

\section{Results and discussion}
\label{sec:results}

\subsection{Waiting time distribution}

\begin{figure}
\includegraphics[bb=0.0cm 0.0cm 8.5cm 6.0cm]{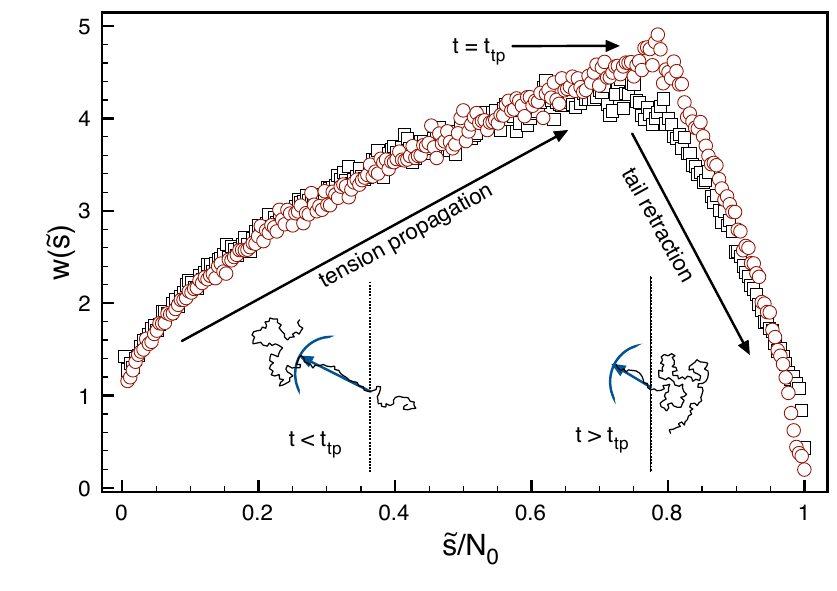}
\caption{(Color online) Comparison of waiting times $w$ for MD (squares) and the BDTP model (circles). The agreement of the BDTP model with MD simulations is excellent, and reveals the two stages of translocation: the tension propagation stage of increasing $w(\tilde{s})$ and the tail retraction stage characterized by decreasing $w(\tilde{s})$. For both MD and BDTP the data has been averaged over 2000 successful translocation events and the system parameters are the same ($N_0=256$, $f=5$, $k_BT=1.2$, $\eta=0.7$).}
\label{fig:wt128_3D}
\end{figure}

We begin the analysis of the results by looking at the waiting time $w(\tilde{s})$ of individual monomers. We do this by solving the BDTP model with parameters typical for molecular dynamics (MD) simulations: $f=5.0$, $k_BT=1.2$, $\eta=0.7$, $\nu=0.588$ (3D) and pore friction $\eta_p=5.0$, corresponding to the pore geometry used in, e.g., Ref.~\cite{luo2009}. The resulting waiting time distribution is an important measure of translocation dynamics. Previously, we have shown that the waiting time distribution is reproduced almost exactly by the BDTP model as compared to MD simulations in 2D and 3D~\cite{ikonen2011}. In addition, the shape of the waiting time distribution is non-monotonic, with the initial part of increasing waiting time  $w(\tilde{s})$ describing the tension propagation stage, and the second part of decreasing  $w(\tilde{s})$ being the tail retraction stage (cf. Fig.~\ref{fig:wt128_3D}). It is reassuring to note that the normalized waiting time distribution for $N_0=256$ looks very similar to that for $N_0=128$ for the same parameters reported earlier in Ref.~\cite{ikonen2011}.

\begin{figure}
\includegraphics[bb=0.0cm 0.0cm 8.5cm 6.0cm]{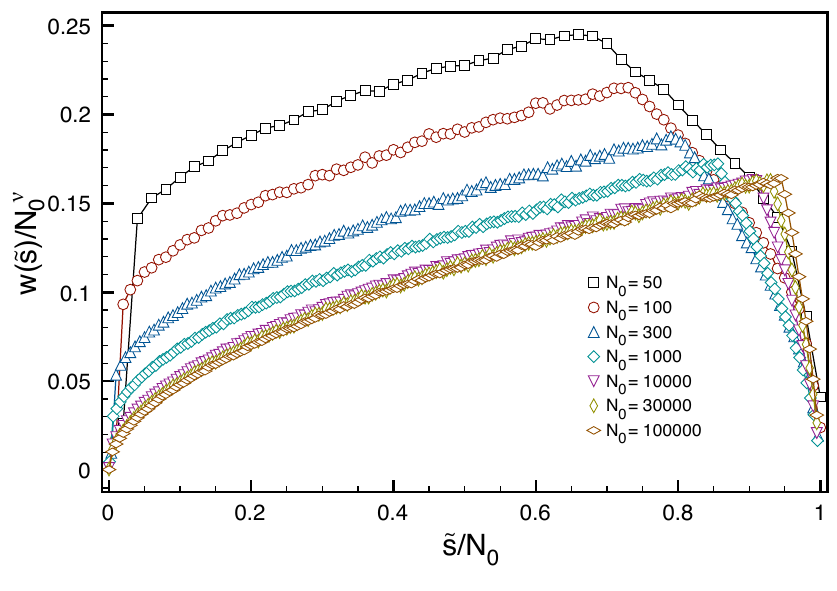}
\caption{(Color online) The waiting times per monomer $w(\tilde{s})$ as a function of monomer number $\tilde{s}$. In the long chain length limit, $w(\tilde{s}/N_0)\sim N_0^{\nu}$, implying the scaling $\tau\sim N_0^{1+\nu}$. Model parameters used are $f=5.0$, $k_BT=1.2$, $\eta=0.7$, $\nu=0.588$ (3D).}
\label{fig:wt_collapse}
\end{figure}

Between the two stages is the maximum of $w(\tilde{s})$, i.e., the moment of maximum friction $\tilde{\Gamma}$, which occurs when the tension front reaches the $N_0$:th monomer of the chain. For sufficiently large $N_0$, the translocation velocity is, according to Eqs.~(\ref{eq:v0_tr}) and~(\ref{eq:v0_sf}), $\tilde{v}_0(\tilde{t}) \propto \tilde{R}(\tilde{t})^{-1}$. Immediately after the tension propagation stage, the location of the tension front is $\tilde{R}\propto N_0^\nu$. Therefore, the maximum waiting time, $w_\mathrm{max}$, should scale with chain length as $w_\mathrm{max}\propto \left[ \tilde{v}_0(\tilde{t}_\mathrm{tp}) \right]^{-1} \propto N_0^\nu$. This is indeed the case, as shown in Fig.~\ref{fig:wt_collapse}, which displays the collapse of the waiting time distributions for different $N_0$ onto a single master curve. Since the area under the $w(\tilde{s})$ curve gives the average translocation time, one has $\tau\sim N_0^\alpha$ with $\alpha\approx{1+\nu}$. However, even for $N_0=10^5$ , the location of $w_\mathrm{max}$ slowly moves to the right and the collapse to the master curve is not exact. This shows that the chain length $N_0=10^5$ is still not in the asymptotic limit! Consequently, for $N_0=10^5$, the scaling of the  average translocation time is also not exactly $\tau \sim N_0^{1+\nu}$, as we will discuss below.

\subsection{Dependence of $\tau$ on the chain length $N_0$}

\begin{figure}
\includegraphics[bb=0.0cm 0.0cm 8.5cm 5.2cm]{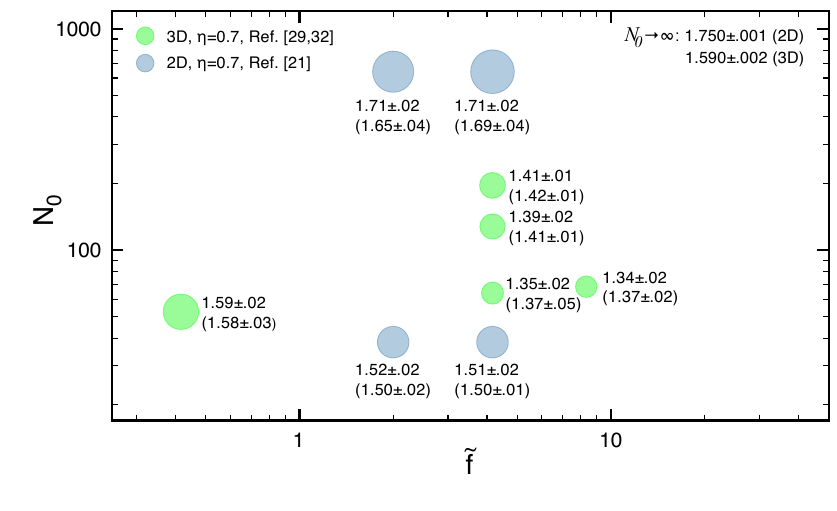}
\caption{(Color online) The exponent $\alpha$ ($\tau\sim N_0^\alpha$) as a function of driving force $\tilde{f}$ and chain length $N_0$. The position of the circle denotes the combination of $(\tilde{f},N_0)$ and, the size reflects the value of  $\alpha$. Next to the symbols, the numerical values of $\alpha$ from the BDTP model are shown in comparison with the values from MD simulations written in parentheses. The asymptotic value of $\alpha$ is indicated in the upper right corner.}
\label{fig:alphas}
\end{figure}

Previously the BDTP model has been compared with MD simulations in several regimes, and has been shown to reproduce the correct  dependence of the average translocation time $\tau$ on the chain length $N_0$ within the numerical accuracy~\cite{ikonen2011}. Some of the results are gathered in Fig.~\ref{fig:alphas}, where we show the translocation time exponent $\alpha$ (defined via $\tau\sim N_0^\alpha$) for both the BDTP model and MD simulations. The figure shows that merely by using the same numerical values for the parameters in the BDTP model and in the MD simulations, the MD results are reproduced by the theory with good accuracy. It is also clear that the numerical value of $\alpha$ depends on several parameters such as chain length, driving force, friction and pore structure. However, as shown by our analysis of the waiting time distribution $w(\tilde{s})$ and discussed in Ref.~\cite{ikonen2011}, all the results shown in Fig.~\ref{fig:alphas} are in the finite chain length regime. Therefore, the apparent non-universality of $\alpha$ is not surprising. To further study the finite size effects, we look at the dependence of $\tau$ on the chain length $N_0$, first for different driving forces $\tilde{f}$ and then for different pore geometries by varying the dimensionless pore friction $\tilde{\eta}_p$.

\subsubsection{Effect of the driving force}

\begin{figure}
\includegraphics[bb=0.0cm 0.0cm 8.5cm 6.0cm]{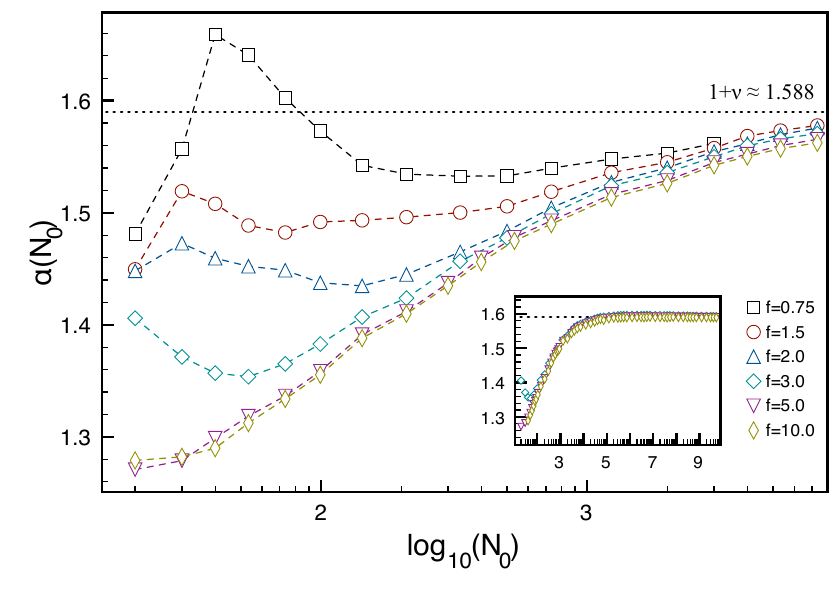}
\caption{(Color online) The effective exponent $\alpha(N_0)\equiv\frac{d\ln \tau}{d\ln N_0}$ as a function of chain length $N_0$ for the BDTP model solved for driving forces $f=0.75,1.5,2.0,3.0,5.0,10.0$. Other parameters are $k_BT=1.2$, $\eta=0.7$, $\eta_p=5.0$. Errors are of the order of the symbol size. Inset: $\alpha(N_0)$ for $f=3.0,5.0,10.0$ up to $N_0=10^{10}$,  showing the approach to the asymptotic value $\alpha(N_0\rightarrow \infty) = 1+\nu$.  }
\label{fig:alpha_different_forces}
\end{figure}

To study the dependence of the average translocation time $\tau$ on the chain length $N_0$, we define the effective (running) exponent  $\alpha(N_0)\equiv\frac{d\ln \tau}{d\ln N_0}$. In the finite chain length regime, the effective $\alpha(N_0)$ is a function of the chain length and approaches the asymptotic value for long chains. In Fig.~\ref{fig:alpha_different_forces}, we show $\alpha(N_0)$ for different driving forces up to chain length $N_0=10^4$. Throughout this regime, the exponent $\alpha(N_0)$ shows clear dependence not only on the chain length, but also on the driving force $f$. This is a clear indication of non-equilibrium behavior and finite chain length effects. As the chain length is increased further, the different curves approach the same asymptotic value of $\alpha(N_0\rightarrow\infty)=1+\nu$, as shown in the inset of Fig.~\ref{fig:alpha_different_forces}. However, the approach is extremely slow: within the numerical accuracy of the BDTP model, the asymptotic value is not reached until $N_0\approx 10^9$.

Another interesting fact is the dependence of $\alpha$ on the driving force $f$. Our results show that as $f$ is increased, $\alpha$ decreases for any fixed chain length $N_0\lesssim 10^4$. However, in the literature, there are conflicting reports on the dependence of $\alpha$ on $f$. In Ref.~\cite{luo2009} it is reported that $\alpha$ decreases as $f$ is increased, in agreement with our results. The BDTP model also agrees with the numerical values of $\alpha$ reported in Ref.~\cite{luo2009} with excellent accuracy (see Fig.~\ref{fig:alphas} and Ref.~\cite{ikonen2011}). On the other hand, MD simulation studies by Lehtola et al.~\cite{lehtola2008} and more recently by Dubbeldam et al.~\cite{dubbeldam2011} report the exactly opposite behavior that $\alpha$ increases with $f$. However, we argue that neither of these studies corresponds to the canonical driven translocation problem, where the chain starts initially on the {\it cis} side and may either successfully translocate or slip back to the {\it cis} side, depending on the fluctuations. In Ref.~\cite{lehtola2008}, the low-force simulations were performed by placing the polymer chain initially halfway through the pore to facilitate successful translocations~\cite{linna_private}. Because of the intrinsic non-equilibrium nature of driven translocation, this approach does not give results that can be directly compared with theory or experiments, where the process always starts with the whole chain initially on the {\it cis} side. In Ref.~\cite{dubbeldam2011}, another method was used to make the low-force simulations possible. Here, the authors prevented the chain from escaping back to the {\it cis} side by making the first monomer too large to fit through the pore. This is equivalent to enforcing an artificial reflecting boundary condition on the first monomer that prevents the escape. Although such a boundary condition has been used several times in the literature, it fundamentally changes the system's behavior in the low-force limit, as we will discuss below.

\subsubsection{Effect of pore size and pore friction $\eta_p$}

\begin{figure}
\includegraphics[bb=0.0cm 0.0cm 8.5cm 6.0cm]{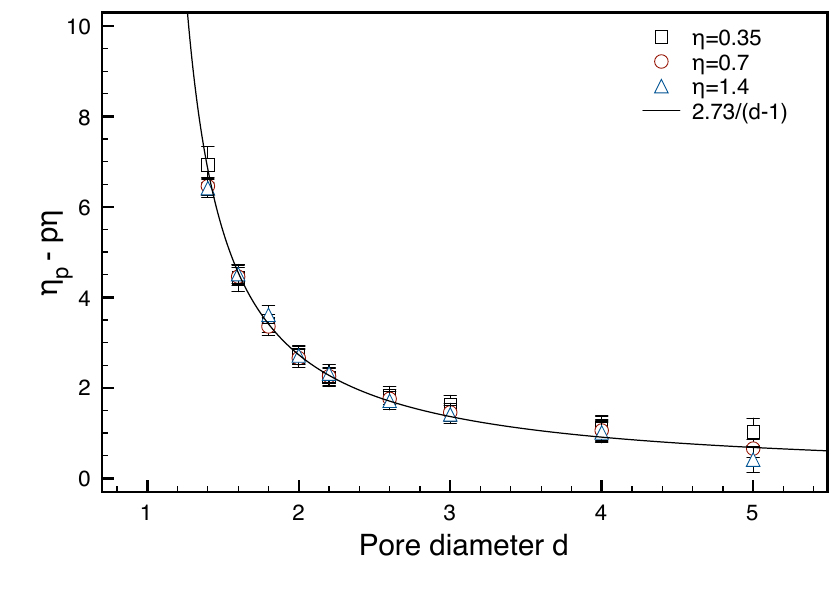}
\caption{(Color online) The pore friction $\eta_p$ as a function of the pore diameter $d$. The symbols indicate the results obtained from MD simulations for different solvent frictions $\eta$, while the solid line is an empirical fitting function. The pore friction $\eta_p$ includes a non-vanishing contribution $p\eta$ from the $p$ monomers inside and in the immediate vicinity of the pore. For $p$, we find $p\approx 2.5$, giving the total pore friction $\eta_p\approx 2.73/(\tilde{d}-1)\eta_\mathrm{LJ}+2.5\eta$, where $\eta_\mathrm{LJ}$ is given by the choice of the Lennard-Jones units (see text). The chain length is $N_0=100$, with other parameters the same as in Fig.~\ref{fig:wt_collapse}.}
\label{fig:pore_friction}
\end{figure}

\begin{figure}
\includegraphics[bb=0.0cm 0.0cm 8.5cm 6.0cm]{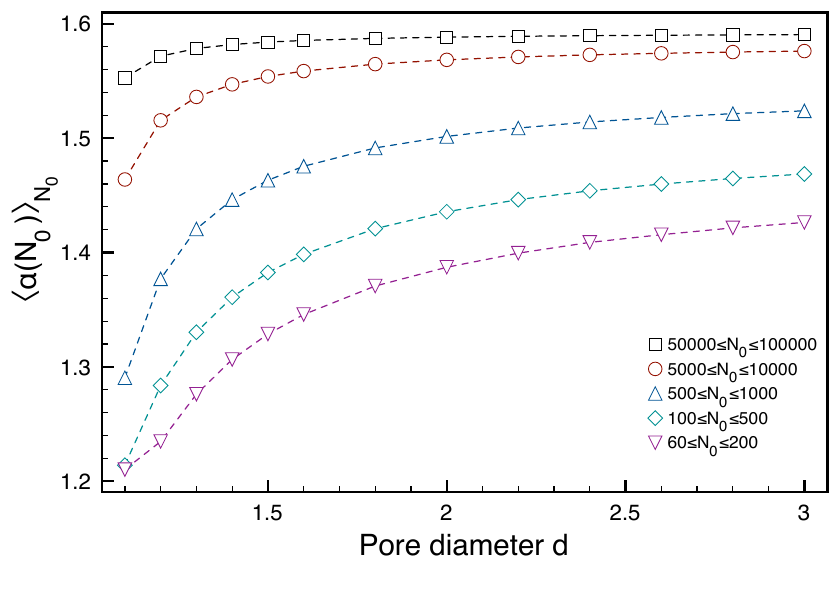}
\caption{(Color online) The effective exponent  $\alpha(N_0)$ averaged over different chain length regimes shown as a function of pore diameter $d$. Model parameters are the same as in Fig.~\ref{fig:pore_friction}.}
\label{fig:alpha_pore_size}
\end{figure}

In the finite chain length regime, it is conceivable that the translocation dynamics is affected by the local neighborhood of the pore. The effect of the pore size and geometry has been previously studied with MD simulations in, e.g., Refs.~\cite{lehtola2010, bhatta2010b, shin2012}. It has been shown that the exponent $\alpha$ depends on the details of the pore, although no systematic study on the nature of the dependence has been performed. In the BDTP model, the effect of pore geometry is mapped into one parameter, the pore friction $\eta_p$. To study the effect of local pore geometry, we have measured $\eta_p$ for different pore diameters $d$ by running MD simulations (for details, see Appendix A) and matching the waiting time distribution with the BDTP model by fixing $\eta_p$ separately for each pore diameter $d$. The results are shown in Fig.~\ref{fig:pore_friction}. For the smaller pore, the interactions between the pore and the polymer are stronger, with $\eta_p$ increasing rapidly as the diameter of the pore approaches the size of the monomer. In addition to the friction between the pore walls and the polymer, $\eta_p$ includes the effective drag force of the monomers inside the pore and in its immediate vicinity. The latter constitutes a non-vanishing contribution to the pore friction, and is extracted from Fig.~\ref{fig:pore_friction} by collapsing the data for different $\eta$ on a single master curve. Empirically, the total pore friction is thus approximately given by $\eta_p\approx 2.73/(\tilde{d}-1)\eta_\mathrm{LJ}+2.5\eta$, where the first term describes the interaction between the pore walls and the polymer, and is similar in form to the one used by Storm et al.~\cite{storm2005}. The factor $\eta_\mathrm{LJ}$ is given by the energy ($\epsilon$) and length scale ($\sigma$) of the Lennard-Jones interaction and the monomer mass $m$ (see Appendix A) as $\eta=\sqrt{m \epsilon / \sigma^2}$. The numerical values may depend on the specific geometry (triangular lattice vs. square lattice, pore length $l_p$, etc.).

In Fig.~\ref{fig:alpha_pore_size}, we examine the dependence of the translocation time exponent $\alpha$ on the pore diameter. Since the exponent $\alpha$ depends on the chain length, we average $\alpha(N_0)$ over different chain length ranges, giving the average $\langle \alpha(N_0)\rangle_{N_0}$ as a function of pore diameter $d$. For extremely long chains, the size of the pore affects the translocation dynamics only slightly. This is because for long chains, the overall friction of the system is dominated by the friction between {\it cis} side subchain and the solvent, with the pore friction adding only a minor contribution. For shorter chains, however, the exponent $\alpha$ clearly decreases with decreasing pore size. For short chains and narrow pores, the pore friction dominates the solvent friction, and the dynamics approaches the constant-friction limit, where $\alpha=1$~\cite{muthu1999}. Similar behavior was seen using MD simulations in Ref.~\cite{bhatta2010b}, where $\alpha$ was measured as 1.35, 1.30 and 1.21 for $64\leq N_0 \leq 256$ and pore diameters 1.5, 1.3 and 1.1, respectively. Using the same parameter values for the BDTP model, we obtain the values 1.37, 1.32 and 1.24 for $\alpha$. The values agree within the statistical error, although there seems to be a systematic error of about 0.02. This may be due to a slightly different pore geometry (triangular vs. our circular) used in Ref.~\cite{bhatta2010b}.

\begin{figure}
\includegraphics[bb=0.0cm 0.0cm 8.5cm 6.0cm]{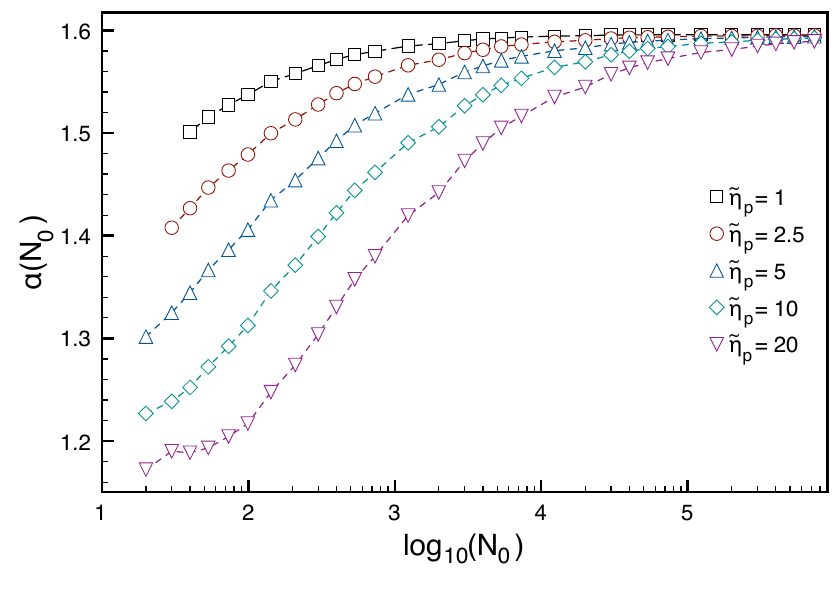}
\caption{(Color online) The effective exponent  $\alpha(N_0)$ as a function of chain length $N_0$ for the BDTP model solved for different ratios $\tilde{\eta}_p$ of pore and solvent friction. Parameters used are the same as in Fig.~\ref{fig:pore_friction}.}
\label{fig:alpha_gp}
\end{figure}

To further illustrate the finite size effect of the pore friction, the effective exponent $\alpha(N_0)$ is solved for different values of the pore friction $\tilde{\eta}_p$. The results are shown in  Fig.~\ref{fig:alpha_gp}. The exponent $\alpha(N_0)$ asymptotically approaches $1+\nu$ for all $\tilde{\eta}_p$. However, the values for finite $N_0$ and the rate of the approach depend on $\tilde{\eta}_p$.  For small $\tilde{\eta}_p$, i.e., wide pores or very viscous solvents, $\alpha(N_0)$ is larger and the asymptotic limit is reached for shorter chains than for large $\tilde{\eta}_p$ (tight pores or low solvent viscosities). Similar results were obtained in the MD simulations of Ref.~\cite{huopaniemi2007}. The data of Fig.~\ref{fig:alpha_gp} indicates that for typical parameters and in the chain length regime relevant for simulations and experiments, the pore friction $\tilde{\eta}_p$ can be a major contribution to the total friction and significantly changes the translocation dynamics.

\subsection{Dependence of $\tau$ on the driving force $f$}

In addition to the exponent $\alpha$, another important measure of translocation dynamics is the dependence of $\tau$ on the driving force $f$. Usually, this dependence is characterized by the scaling exponent $\delta$, defined as $\tau\sim f^\delta$. The simplest argument, namely translocation at constant velocity $\tilde{v}_0\sim \tilde{f}$, gives the scaling exponent $\delta = -1$, which is also supported by some theoretical studies~\cite{sung1996,muthu1999,kantor2004,rowghanian2011,dubbeldam2011}, numerical simulations~\cite{huopaniemi2006,lehtola2008,lehtola2009,luo2008,bhatta2009,bhatta2010} and experiments~\cite{kasi1996}. For the deterministic tension propagation theory, Sakaue predicts that $\delta=-1$ for the SS and equilibrium regimes, $\delta\approx -1.26$ for the TR regime and $\delta \approx -0.74$ for the SF regime. In Refs.~\cite{rowghanian2011,dubbeldam2011}, on the other hand, the exponent $\delta = -1$ is predicted even for the TR and SF regimes. In most MD simulations, the measured exponent is slightly larger than $-1$, typically between $\delta \approx -0.9$ and $\delta \approx -0.97$. For the BDTP model, we have measured $\delta\approx -0.94$ for $N_0=128$, $k_BT=1.2$, $\eta=0.7$ and $0.25\leq f \leq 15$, which is in good agreement with both experiments and simulations.

\begin{figure}
\includegraphics[bb=0.0cm 0.0cm 8.5cm 6.0cm]{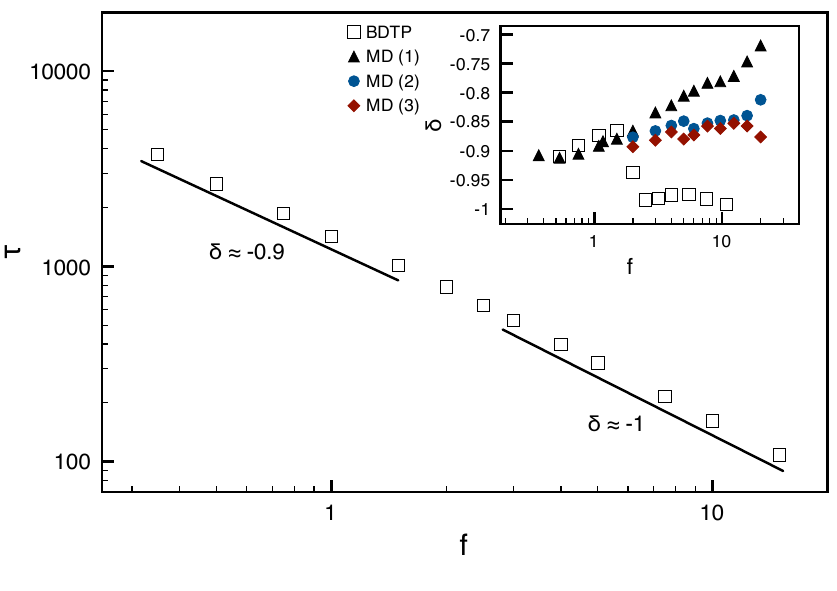}
\caption{(Color online) The dependence of $\tau$ on $f$ measured from the BDTP model with $N_0=128$ and $k_BT=1.2$, $\gamma=0.7$. The inset shows the exponent $\delta$ ($\tau\sim f^\delta$) as a function of $f$ for both BDTP and MD. Here MD (1) corresponds to $\eta=0.7$ and $k=15$, MD (2) to $\eta=10$ and $k=15$ and MD (3) to $\eta=10$ and $k=150$. For $\delta$, the error is of the order of the size of the symbols. }
\label{fig:delta_BDTP}
\end{figure}

However, looking at $\delta$ more closely reveals interesting details. As shown in Fig.~\ref{fig:delta_BDTP}, for $f  \lesssim 1$, $\delta\approx -0.9$ and turns over to $\delta \approx -1$ for $f \gg 1$. Similar behavior for the low-force regime was reported in~\cite{luo2009}. However, in the high force regime the MD simulations of Ref.~~\cite{luo2009} give $\delta\approx -0.8$ instead of $\delta\approx -1$. On the other hand, the scaling $\delta \approx -1$ given by BDTP for large $f$ is in agreement with the theoretical prediction of Refs.~\cite{sakaue2007,sakaue2008,sakaue2010,saito2011,rowghanian2011,dubbeldam2011}. To investigate the matter more closely, we have performed extensive MD simulations. We have used the same parameters ($\eta=0.7$, $k_BT=1.2$) as in Ref.~\cite{luo2009} to allow direct comparison. In addition, we have studied the effect of solvent viscosity by running simulations with $\eta=10.0$ and the effect of bond strength by using a FENE spring constant $k=150$ (hard bonds) and $k=15$ (soft bonds).

The results of the MD simulations are shown in the inset of Fig.~\ref{fig:delta_BDTP}. The effective exponent $\delta$ as a function of the force is found by linear least squares fit from three consecutive [$\log(\tau),\log(f)$] data points. For low to intermediate forces, we measure $\delta \approx -0.9$, in agreement with Ref.~\cite{luo2009} and the BDTP model. For the large forces, we have $-0.8 \lesssim \delta \lesssim -0.7$ for $\eta=0.7$ and $k=15$, also in agreement with Ref.~\cite{luo2009}. However, for the large driving forces the MD results depend on the friction $\eta$ and the spring constant $k$. For the smaller $\eta$, one has significantly larger $\delta$, and similarly for the spring constant $k$. The reason for the former is that for very low friction, the response of the system to the force is not linear due to the inertial term in the equations of motion (see Appendix A). When the friction is increased, the mass term becomes less significant and the exponent $\delta$ decreases. Similarly, for extremely high forces and small $k$, the bonds can be significantly stretched by the driving force. This increases the exponent $\delta$ for small $k$, as shown in Fig.~\ref{fig:delta_BDTP}. However, we have confirmed that even using overdamped dynamics without the inertial term (see Appendix A) and sufficiently hard bonds, the exponent does not reach $-1$, but stays between$ -0.9 \lesssim \delta\lesssim -0.95$. Therefore, although the low friction and soft bonds typically used in MD simulations contribute significantly to the difference between BDTP and MD, they do not explain it fully. The most probable cause for the remaining difference is the absence of the {\it trans} side subchain from the BDTP model. For sufficiently high $f$, significant crowding of monomers close to the pore on the {\it trans} side occurs, which could increase the scaling exponent $\delta$. For very long chains, the effect should be small, because the high friction due to the long tail on the {\it cis} side leads to slow translocation even for large $f$, and the friction due to crowding becomes less significant.

\subsection{Time evolution of the translocation coordinate $s$}

\begin{figure}
\includegraphics[bb=0.0cm 0.0cm 8.5cm 6.0cm]{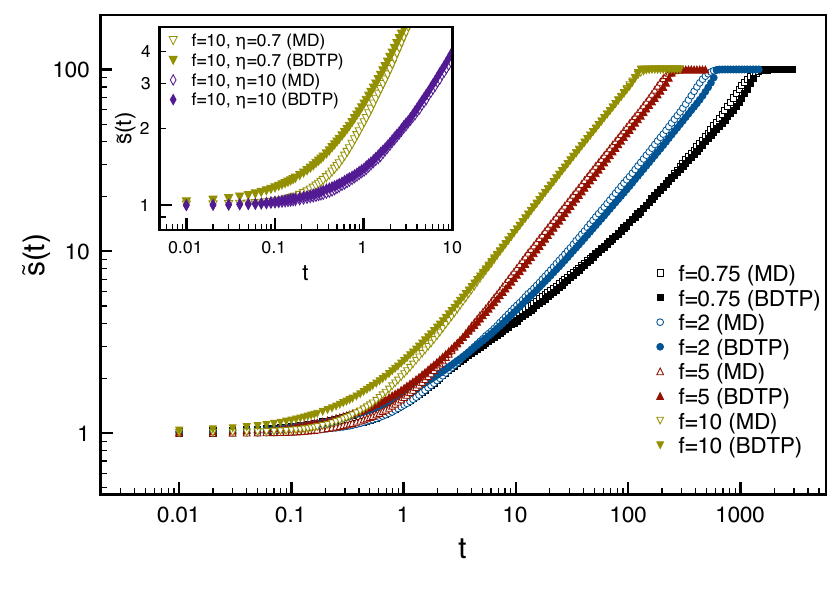}
\caption{(Color online) The translocation coordinate $\tilde{s}$ as a function of time for the BDTP model (filled symbols) and MD simulations (open symbols) for $N_0=100$. Other parameters are the same as in Fig.~\ref{fig:wt_collapse}. Inset: $\tilde{s}(t)$ for $\eta=0.7$ and $\eta=10$, showing the difference between the inertial and overdamped regimes.}
\label{fig:s_BDTP_MD}
\end{figure}

Finally, we look at the time evolution of the translocation coordinate $\tilde{s}(t)$. In Fig.~\ref{fig:s_BDTP_MD} we compare the results given by the BDTP model with MD simulations for a fixed chain length $N_0=100$ with different driving forces. Overall, the BDTP model agrees very well with the MD results. The curves practically overlap, except for the beginning of the process, where the $\tilde{s}$ measured from MD simulations lags behind the BDTP solution. This difference increases with the driving force $f$. However, as shown in the inset of Fig.~\ref{fig:s_BDTP_MD}, the difference disappears when the solvent friction is increased from $\eta=0.7$ to $\eta=10.0$. This shows that the difference is caused by the inertial effects in the low-friction MD simulations: here the monomers of mass $m$ need to be accelerated for a time $t_\mathrm{acc}\propto m/\eta$ before they reach the friction-limited velocity.

\begin{figure}
\includegraphics[bb=0.0cm 0.0cm 8.5cm 5.2cm]{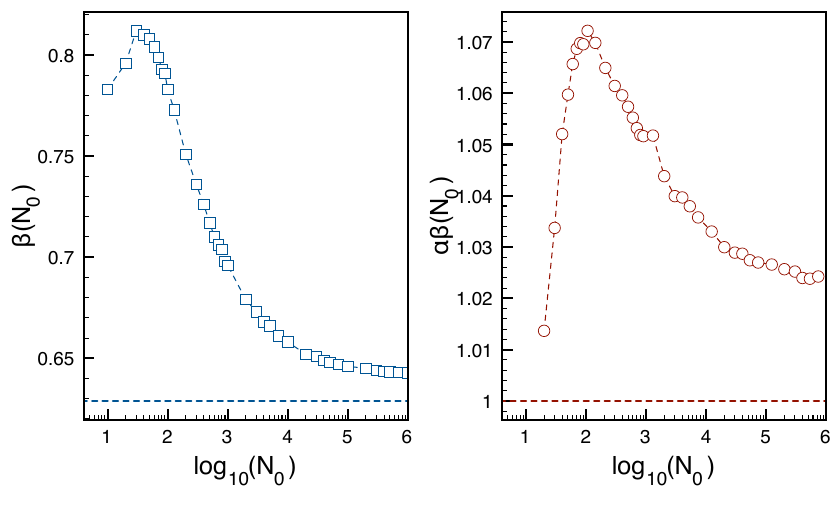}
\caption{(Color online) The exponent $\beta$ ($\langle \tilde{s}(t)\rangle\sim t^\beta$, left panel) and the product of the exponents $\alpha$ and $\beta$ (right panel) as a function of chain length. The dashed lines indicate the theoretical asymptotic values $\beta(N_0\rightarrow\infty)=1/(1+\nu)$ and $\alpha\beta(N_0\rightarrow\infty)=1$. Model parameters are the same as in Fig.~\ref{fig:s_BDTP_MD}. }
\label{fig:alpha_beta}
\end{figure}

The time-dependence of $\tilde{s}$ can also be characterized by the scaling exponent $\beta$, which we define as $\langle \tilde{s}(t) \rangle \sim t^{\beta}$\footnote{In the literature, $\beta$ is sometimes defined through $\langle \tilde{s}^2(t) \rangle \sim t^{\beta}$. However, our conclusions remain the same, irrespective of the definition.}. For the chain lengths $100 \leq N_0 \leq 500$ we obtain $\beta=0.75$, which agrees with the MD simulations of Ref.~\cite{luo2008} within the statistical accuracy, and is also similar to the value reported in Ref.~\cite{bhatta2009}, where slightly different parameter values were used. We have also solved $\beta$ from the BDTP model for much longer chains. The results are shown in Fig.~\ref{fig:alpha_beta} for $N_0\leq 10^6$. Once again, the data shows the extremely slow approach to the asymptotic limit. Even for $N_0=10^{10}$, the value of $\beta$ continues to decrease, although extremely slowly (not shown).

In addition to $\beta$, we examine the product  $\alpha \beta$. Since at the moment of translocation, $\tilde{s}(\tau)=N_0$, the exponents $\alpha$ and $\beta$ are related by $\alpha\beta=1$. The relation can be exact only in the asymptotic limit $N_0\rightarrow\infty$, since for finite chain lengths, the scaling $\tilde{s}(t)\sim t^\beta$ is not exact, as shown in Fig.~\ref{fig:s_BDTP_MD}. However, we expect the relation to hold approximately even for finite $N_0$. The results are shown in Fig.~\ref{fig:alpha_beta}. For short chains, the product quickly increases from $\alpha\beta\approx 1$ to $\alpha\beta\approx 1.07$ at $N_0\approx 100$, where it attains its maximum value. Thereafter the product slowly decreases, approaching the theoretical asymptotic limit $\alpha\beta = 1$. Also in this case, the approach is extremely slow. For $N_0=10^{10}$ we have measured $\alpha\beta\approx 1.015$, with the value still gradually decreasing towards 1. In the finite chain length regime, the BDTP model is in excellent agreement with the available MD simulation data. For $100 \leq N_0 \leq 500$ we have on the average $\alpha\beta = 1.065$, which matches exactly with the results of Ref.~\cite{luo2008}.

\begin{figure}
\includegraphics[bb=0.0cm 0.0cm 8.5cm 6.2cm]{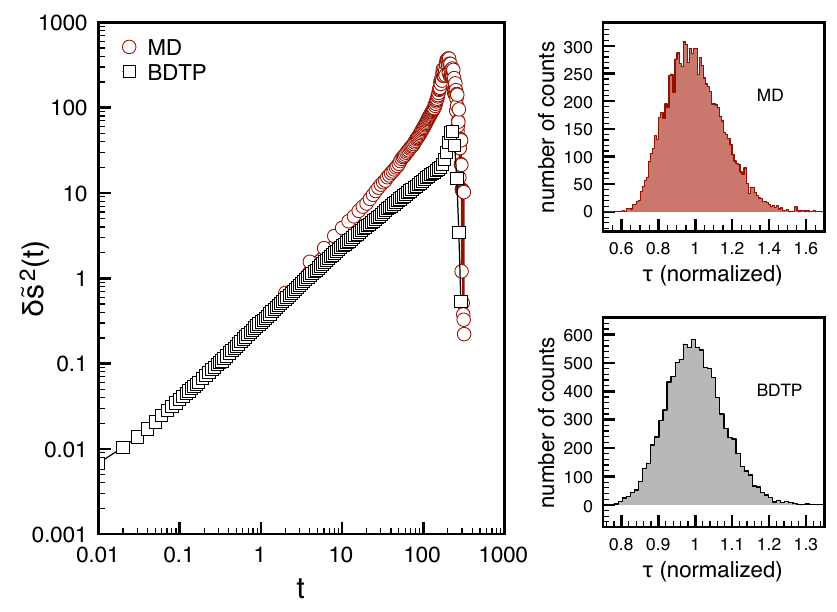}
\caption{(Color online) The fluctuations of the translocation coordinate $\tilde{s}$ (left) and the distribution of translocation times (right) for the BDTP model and molecular dynamics simulations. In the BDTP model, the {\it trans} side subchain is not modeled, which reduces fluctuations.}
\label{fig: delta_s_hist_BDTP_MD}
\end{figure}

Finally, we look at the fluctuations of the BDTP model. To do this, we have measured the fluctuations of the translocation coordinate $\langle \delta \tilde{s}^2(t) \rangle \equiv \langle \tilde{s}^2(t) \rangle  -\langle \tilde{s}(t) \rangle ^2$ as a function of time for both the BDTP model and directly from MD simulations. In the BDTP model, the {\it trans} side subchain is not included, and therefore we expect that the total amount of noise is smaller in the BDTP model than in MD. This is indeed the case, as is shown in Fig.~\ref{fig: delta_s_hist_BDTP_MD}. Initially, $\langle \delta \tilde{s}^2(t) \rangle$ scales similarly for the BDTP and MD, but as the chain translocates and the length of the chain on the {\it trans} side increases, the fluctuations in the MD simulation start to increase faster. The effect of the missing {\it trans} side fluctuations can be also seen in the distribution of the translocation times. As shown in Fig.~\ref{fig: delta_s_hist_BDTP_MD}, the standard deviation of the  translocation times is roughly two times larger in the MD simulations, as compared to the BDTP solutions. Including additional noise from the translocated part of the chain may bring both the scaling of $\langle \delta \tilde{s}^2(t) \rangle$ and the distribution of translocation times closer to the MD results. However, a proper implementation of the {\it trans} side fluctuations would also require considering the out-of equilibrium effects of monomer crowding in front of the {\it trans} side pore entrance. This is a subject of future study and is not within the scope of the present work.

\subsection{Initial stages of translocation}

Lastly, we wish to discuss the initial stages of the translocation process. Several authors have suggested that in the early stages of translocation, an initial tension blob should be formed at the pore entrance before the tension propagation process can begin~\cite{rowghanian2011, saito2011, dubbeldam2011}. The theory predicts that the time it takes for the initial blob to form, $\tau_\mathrm{init}$, decreases with increasing force, as the size of the blob gets smaller. In MD simulations, this process should be visible as an initial period of time during which the translocation coordinate $\tilde{s}$ changes only slightly (say, from 1 to 2).

\begin{figure}
\includegraphics[bb=0.0cm 0.0cm 8.5cm 6.0cm]{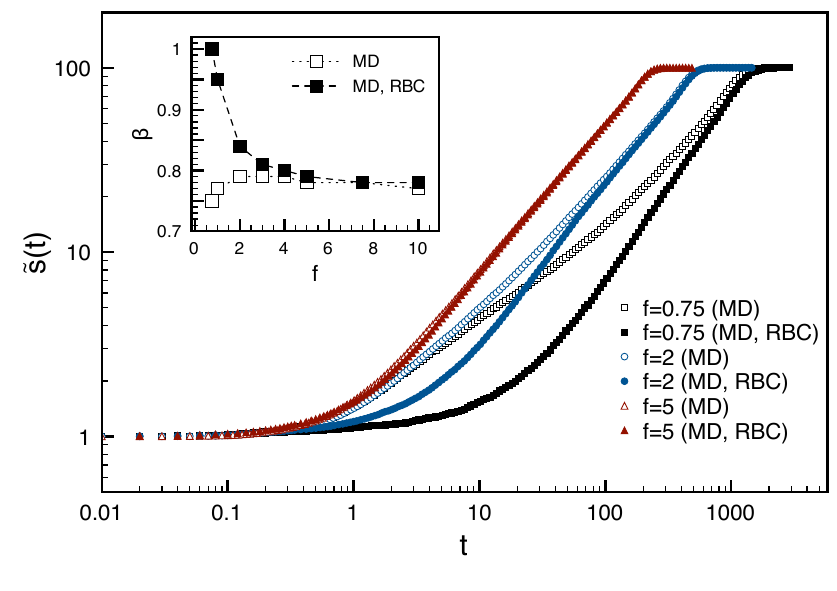}
\caption{(Color online) The translocation coordinate $\tilde{s}$ as a function of time from MD simulations. The chain is either allowed to escape the pore to the {\it cis} side (open symbols, same data as in Fig.
 \ref{fig:s_BDTP_MD}) or such an escape is prevented by a reflecting boundary condition (RBC, filled symbols). The results change drastically, if the RBC is used. Inset: the scaling exponent $\beta$ ($\langle \tilde{s}(t)\rangle\sim t^\beta$) as a function of the driving force $f$ both without and with the RBC. Parameters are the same as in Fig.~\ref{fig:s_BDTP_MD}}
\label{fig:s_MD_boundary}
\end{figure}

We have investigated the prediction by running extensive MD simulations and looking at the time evolution of $\tilde{s}$ and the dependence of the corresponding scaling exponent $\beta$ on the driving force. The results are shown in Fig.~\ref{fig:s_MD_boundary} as open symbols. At least for $\tilde{f} \gtrsim 0.6$, the initial time evolution of $\tilde{s}$ seems to be insensitive to the driving force. Hence, we find no indications of the blob initiation process. In addition, we find that the scaling exponent $\beta$, as measured using values in the range $N_0/2 \lesssim \tilde{s} \lesssim N_0 $, is also quite insensitive to the driving force for $\tilde{f} \gtrsim 3$, and decreases with decreasing force for $\tilde{f} \lesssim 1$. This result is in agreement with the BDTP model's result that for low forces the exponent $\alpha$ decreases with {\it increasing} force and the approximate relation $\alpha\beta \approx 1$. However, the result is contrasted by the recent study by Dubbeldam et al.~\cite{dubbeldam2011} where it is found that $\beta$ decreases with {\it increasing} force, and where the blob initiation process is observed to increase the time that the system remains close to the initial value of $\tilde{s}\approx 1$.

In an attempt to resolve the contradiction, we have repeated the MD simulations with an additional reflecting boundary condition (RBC) that prevents the escape of the first monomer to the {\it cis} side. A similar boundary condition was also used in the MD simulations of Ref.~\cite{dubbeldam2011}. The results are shown in Fig.~\ref{fig:s_MD_boundary} as solid symbols. In this case, the results agree with Ref.~\cite{dubbeldam2011}, showing a decrease of $\beta$ with increasing force, and correspondingly a decrease in the time that the system remains close to $\tilde{s}\approx 1$. The behavior of the translocation process in these two cases (with or without the RBC) is therefore qualitatively different, especially at low driving forces. 

We argue that the difference is due to a fundamental change in the system's free energy, introduced by imposing the reflective boundary condition. For the canonical translocation problem, where the chain is allowed to escape to the {\it cis} side, the free energy has a maximum but no minimum~\cite{sung1996, muthu1999}. Preventing the escape to the {\it cis} side creates a local minimum in the free energy where the system can oscillate, changing the problem into a thermally activated escape process, similar to the famous Kramers problem~\cite{kramers1940}. Therefore, the initial period of slow growth of $\tilde{s}$ is not related to the blob initiation, but to the thermal motion of the chain in the free-energy well near $\tilde{s}\approx 1$, which precedes the eventual escape across the free-energy barrier to the {\it trans} side. In fact, according to our MD simulations with the RBC, the time that is spent in evolving from $\tilde{s}=1$ to, say, $\tilde{s}=2$, increases roughly exponentially with decreasing force. This is a clear indication of a thermally activated barrier crossing process. In addition, the influence of the RBC disappears for large driving forces where the activation barrier becomes negligible, also in support of our argument.  

Finally, we should acknowledge that while our MD simulations show no signs of the blob initiation process, it is possible that the process becomes practically observable only for significantly smaller driving forces ($\tilde{f}\ll 1$). Unfortunately, that regime may be out of reach of current MD simulations.

\section{Conclusions}

In this work, we have employed theoretical methods and Langevin dynamics simulations to examine the finite chain length effects and the resulting non-universalities in driven polymer translocation. We show that the driven translocation problem can be accurately described by the non-equilibrium tension propagation mechanism proposed by Sakaue~\cite{sakaue2007}. However, finite chain length effects prevail for extremely long chains. Thus, for chain lengths relevant for experiments or numerical simulations, these effects need to be taken into account. Especially, the frictional interaction between the pore and the polymer has to be included in the theory, if quantitative accuracy is required. This result underlines the significance of understanding the interactions between the polymer and the pore also from the point of view of fundamental physics, and is especially important for interpretation of experimental data.

In addition, we discuss the potential limitations of the present tension propagation theory. Although very accurate in the typical experimental and simulation regimes, the absence of the {\it trans} side subchain from the theory may result in less-than-accurate predictions in the extremely strong force regime ($f  \gtrsim k_BTN_0^\nu$) and in the extremely low force regime ($f \lesssim k_BT N_0^{-\nu}$). In the first case, the non-equilibrium crowding of monomers on the {\it trans} side could increase the effective friction. In the other extreme, the fluctuations from the {\it trans} side chain, which are neglected in the model, may become important. In addition, the tension propagation theory of Refs.~\cite{sakaue2007,sakaue2008,sakaue2010,saito2011,dubbeldam2011,rowghanian2011} is inherently a deterministic theory. However, in the low-force regime the tension propagation process becomes increasingly stochastic, instead of deterministic. Exploring the implications of diffusive tension propagation is not within the scope of the present work, but is a subject of future study.

\acknowledgments

This work has been supported in part by the Academy of Finland through its COMP Center of Excellence and Transpoly Consortium grant. TI acknowledges the financial support of the Finnish Doctoral Programme in Computational Sciences (FICS) and the Finnish Foundation for Technology Promotion (TES). AB has been partially supported by the NSF-CHEM grant \#0809821. The authors also wish to thank CSC, the Finnish IT center for science, for allocation of computer resources.

\appendix
\section{Molecular dynamics simulations}

The details of the molecular dynamics simulations that we have used for benchmarking are explained in this appendix. In the MD simulations, the polymer chain is modeled as Lennard-Jones particles interconnected by nonlinear FENE springs. Excluded volume interaction between monomers is given by the short-range repulsive Lennard-Jones potential:
\begin{equation}
U_\mathrm{LJ}(r)=
\begin{cases} 4\epsilon \left[ \left(\frac{\sigma}{r}\right)^{12} -  \left(\frac{\sigma}{r}\right)^{6}\right] +\epsilon & \text{for } r\leq 2^{1/6}\sigma\\
0 & \text{for } r>2^{1/6}\sigma
\end{cases}
\end{equation}
Here, $r$ is the distance between monomers, $\sigma$ is the diameter of the monomer and $\epsilon$ is the depth of the potential well. Consecutive monomers are also connected by FENE springs with 
\begin{equation}
U_\mathrm{FENE}(r)=-\frac{1}{2}kR_0^2\ln (1-r^2/R_0^2),
\end{equation}
where $k$ is the FENE spring constant and $R_0$ is the maximum allowed separation between consecutive monomers. For the chain, we use the parameters $\epsilon=1$, $k=15$ and $R_0=2$, unless otherwise indicated. The main part of the wall is constructed using a repulsive external potential of the Lennard-Jones form $U_\mathrm{ext}=4\epsilon \left[ \left(\frac{\sigma}{x}\right)^{12} -  \left(\frac{\sigma}{x}\right)^{6}\right] +\epsilon$ for $x\leq 2^{1/6}\sigma$ and 0 otherwise. Here $x$ is the coordinate in the direction perpendicular to the wall, with $x<0$ signifying the {\it cis} side and $x>0$ the {\it trans} side. The neighborhood of the pore is constructed of immobile Lennard-Jones beads of size $\sigma$. All monomer-pore particle pairs have the same short-range repulsive LJ-interaction as described above. We have verified that using the simple external potential $U_\mathrm{ext}$ for the wall gives the same results (within statistical error) as using a wall made of monomers in fixed lattice sites, at least as long as the interaction between the wall and the polymer is purely repulsive. On the other hand, the geometry of the pore itself may have significant effect on the results, as is discussed in the main text.

\begin{figure}
\includegraphics[bb=0.0cm 0.0cm 8.5cm 4.2cm]{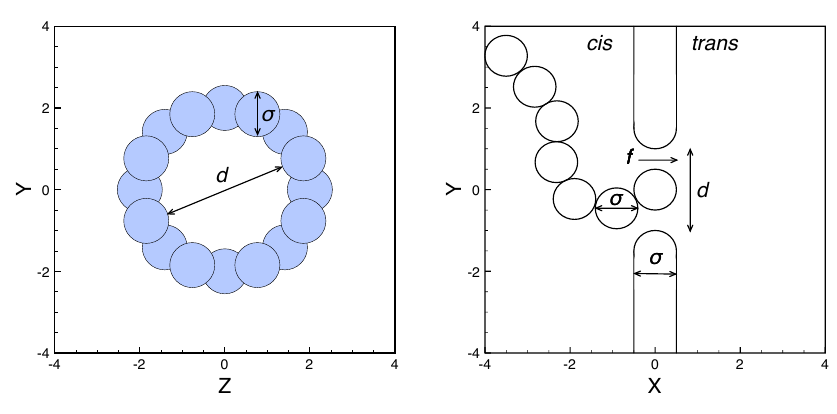}
\caption{(Color online) Left panel: geometry of the pore used in the 3D MD simulations. The pore is formed by placing 16 monomers of diameter $\sigma$ equidistantly on a circle of radius $(d+\sigma)/2$, resulting in a pore of diameter $d$. Right panel: a side-view of the pore and the wall, also showing a typical initial configuration of the chain and the direction of the driving force $f$. }
\label{fig:pore}
\end{figure}

\begin{figure}
\includegraphics[bb=0.0cm 0.0cm 8.5cm 13.2cm]{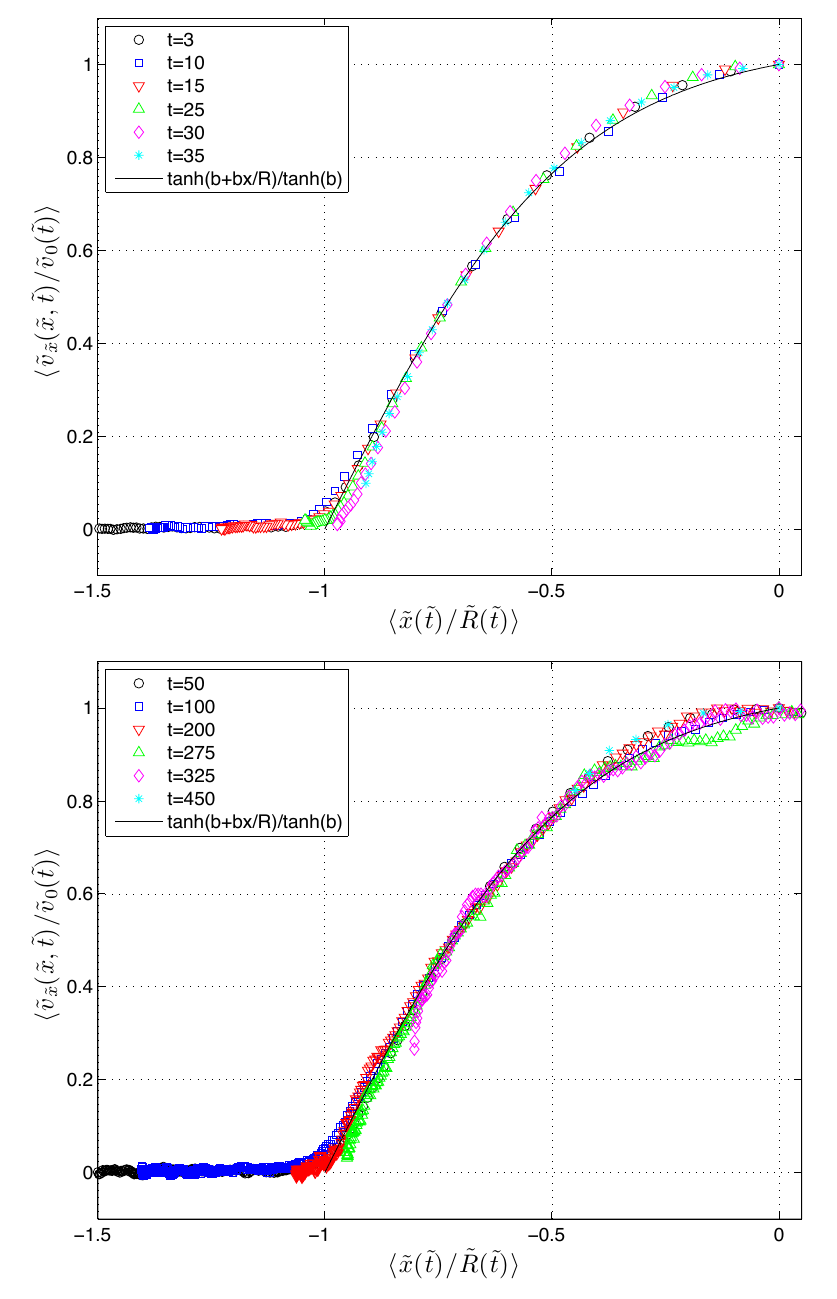}
\caption{(Color online) The velocity perpendicular to the wall of individual monomers $\tilde{v}_x$ for chain lengths $N_0=100$ (upper panel) and $N_0=500$ (lower panel) as a function of normalized perpendicular distance $\tilde{x}$ from the wall. The location of the tension front is given by $\tilde{R}$. The different plots corresponds to different instances in time, with the average $\langle\cdot\rangle$ taken over 10000 independent runs and a time window $\Delta t=1$. The solid black line indicates the empirical fitting function of Eq.~(\ref{eq:v_profile}) with $b\approx 1.9$. In the BDTP model, the parameter $b$ is not found by fitting, but fixed by global mass conservation. However, the resulting numerical value is comparable to the value found in MD simulations. }
\label{fig:v_profile}
\end{figure}

Similarly to most of the molecular dynamics simulations in the literature~\cite{huopaniemi2006,lehtola2008,lehtola2009,bhatta2009,bhatta2010,luo2008,luo2009,metzler2010,dubbeldam2011}, we take the surrounding solvent into account through frictional and random forces. Thus, each monomer is described by the Langevin equation of motion 
\begin{equation}
m \mathbf{\ddot{r}}_i=-\nabla (U_\mathrm{LJ}+U_\mathrm{FENE} + U_\mathrm{ext})-\eta \mathbf{v}_i + {\zeta}_i,\label{eq:motionMD}
\end{equation}
where $m$ is the monomer mass, $\eta$ is the friction coefficient, $\mathbf{v}_i$ is the monomer velocity, $U_\mathrm{ext}$ is the external potential that gives the is the external force $f=-\nabla U_\mathrm{ext}$ in the pore and $\zeta_i$ is the random force with the correlations $\langle  \zeta_i(t) \zeta_j(t') \rangle = 2\eta k_BT \delta_{i,j}\delta(t-t')$, where $k_BT$ is the thermal energy. Typically, we have used the parameter values $m=1$, $\eta=0.7$, $k_BT=1.2$. The equations of motion are solved with the BBK algorithm~\cite{BBK} with time step $\delta t = 0.005$. Initially, the polymer chain is placed at the pore entrance as shown in Fig.~\ref{fig:pore}. Eq.~(\ref{eq:motionMD}) is then solved numerically while keeping the first monomer fixed until an uncorrelated initial configuration is generated. After that, the whole chain is allowed to evolve according to Eq.~(\ref{eq:motionMD}) until the chain escapes either to the {\it cis} or {\it trans} side. The latter is recorded as a successful translocation event. Typically, we average our data over at least $10^4$ such events.

In analyzing the dependence of translocation time $\tau$ on the driving force $f$ we also used overdamped Langevin dynamics to study the translocation dynamics in the limit of negligible inertial effects. In this case, we neglect the inertial term in the equations motion, giving the overdamped Langevin equations 
\begin{equation}
\eta \mathbf{v}_i =-\nabla (U_\mathrm{LJ}+U_\mathrm{FENE} + U_\mathrm{ext})+\zeta_i,\label{eq:motionMDoverdamped}
\end{equation}
for each monomer. The conventions are the same as in Eq.~(\ref{eq:motionMD}). Eqs.~(\ref{eq:motionMDoverdamped}) are solved with the explicit Euler method~\cite{kloeden-platen} with time step $\delta t \leq 0.0001$. In this case, we average our data over $10^5$ successful translocation events.

\section{Velocity profile}

We have measured the velocity profile $\tilde{v}(\tilde{x},\tilde{t})$ of the {\it cis} side subchain by running extensive MD simulations in 3D, with a relatively high driving force $f=10$. In Fig.~\ref{fig:v_profile} we present the velocity profile for $N_0=500$ obtained as an average over 10000 independent translocation events and a short time window of $\Delta t =1$. From the figure, one immediately notices that the curves corresponding to different times $t$ collapse onto a single master curve if the perpendicular distance $\tilde{x}$ from the wall is scaled by the location $\tilde{R}$ of the tension front, defined as the distance where the average velocity goes to zero and, if the velocity is scaled by the maximum velocity. This master curve can be approximated by the expression 
\begin{equation}
\tilde{v}(\tilde{x},\tilde{t})=\tilde{v}_0(\tilde{t}) \frac{\tanh \left[ b\left( \tilde{x}/\tilde{R}+1\right)\right]}{\tanh(b)},\label{eq:v_profile}
\end{equation} where $b$ is the parameter that controls the sharpness of the profile.  In principle, $b$ would be nontrivial function of several parameters, such that $b=b(\nu,f,k_BT,\eta)$. However, in the BDTP model, the parameter $b$ {\it is not found by fitting to MD data, but fixed by enforcing global mass conservation, i.e., requiring internal consistency within the model.} 

In addition, it turns out that for the purposes of the BDTP model, the exact form of the velocity profile is not crucial. The model seems to be very robust with respect to the different forms of the profile, as long as the number of monomers is globally conserved. In fact, we also implemented the model with a piecewise linear velocity profile such that $\tilde{v}(\tilde{x},\tilde{t})=\tilde{v}_0$ for $\tilde{x}\geq -\tilde{R}+\Delta\tilde{R}$ and $\tilde{v}(\tilde{x},\tilde{t})=\tilde{v}_0 \frac{\tilde{x}+\tilde{R}}{\Delta\tilde{R}}$. Here $\Delta R$ is a parameter that controls the shape of the velocity profile and is fixed by requiring conservation of mass, similar to the parameter $b$ in Eq.~\ref{eq:v_profile}. Typically, the difference in numerical results given by the piecewise linear profile and those given by Eq.~(\ref{eq:v_profile}) was comparable to the statistical uncertainty, although Eq.~(\ref{eq:v_profile}) seems to give a slightly better match with MD simulations. Furthermore, in both cases, the asymptotic limit of the exponent $\alpha$ is $\alpha_\infty=1+\nu$.


\begin{thebibliography}{99}
\bibitem{approx} The exact relation would be $\tilde{f}_0 = \tilde{f}_\mathrm{tot}-\tilde{\sigma}_0 \tilde{\eta}_p \tilde{v}_0$. The approximation becomes exact in the SF regime. However, even for the smallest forces considered in the present work, the approximation introduces only a small deviation for short chains, and gives the correct asymptotic ($N\rightarrow\infty$) limit.
\end{thebibliography}
\end{document}